\documentclass[conference]{IEEEtran}
\IEEEoverridecommandlockouts
\usepackage{cite}
\usepackage{amsmath,amssymb,amsfonts}
\usepackage{algorithmic}
\usepackage{graphicx}
\usepackage{textcomp}
\usepackage{xcolor}
\usepackage{booktabs}
\usepackage{bm}

\def\BibTeX{{\rm B\kern-.05em{\sc i\kern-.025em b}\kern-.08em
    T\kern-.1667em\lower.7ex\hbox{E}\kern-.125emX}}
\begin{document}

\title{SUSing: SU-net for Singing Voice Synthesis\\
\thanks{$\ast$Corresponding author: Jianzong Wang (jzwang@188.com).}
}

\author{\IEEEauthorblockN{Xulong Zhang, Jianzong Wang$^{\ast}$, Ning Cheng, Jing Xiao}
\IEEEauthorblockA{\textit{Ping An Technology (Shenzhen) Co., Ltd., China} }
}
\maketitle

\begin{abstract}
Singing voice synthesis is a generative task that involves multi-dimensional control of the singing model, including lyrics, pitch, and duration, and includes the timbre of the singer and singing skills such as vibrato. In this paper, we proposed SU-net for singing voice synthesis named SUSing. Synthesizing singing voice is treated as a translation task between lyrics and music score and spectrum. The lyrics and music score information is encoded into a two-dimensional feature representation through the convolution layer. The two-dimensional feature and its frequency spectrum are mapped to the target spectrum in an autoregressive manner through a SU-net network. Within the SU-net the stripe pooling method is used to replace the alternate global pooling method to learn the vertical frequency relationship in the spectrum and the changes of frequency in the time domain. The experimental results on the public dataset Kiritan show that the proposed method can synthesize more natural singing voices.
\end{abstract}

\begin{IEEEkeywords}
Singing voice synthesis, SU-net, Stripe pooling, Convolution
\end{IEEEkeywords}

\section{Introduction}

Singing voice synthesis (SVS) generates a singing sound with musical characteristics according to a given lyrics or music score. The task of singing voice synthesis is similar to the text-to-speech (TTS) in speech processing, and the synthesis speech is generated according to the given text. With the development of text-to-speech technology, many technologies~\cite{yi2019singing,zhang2020research,chen2020hifisinger,hono2021sinsy,qubo2021,xue2021learn2sing,gao2021vocal} have been successfully applied to the task of singing voice synthesis. Both of the tasks of TTS and SVS encoded the lyrics or text into an acoustic variable, through a vocoder to synthesize the audio waveform. Kong \textit{et al.}~\cite{kong2020hifi} proposed HiFi-GAN, which contains a discriminator which consists of small sub-discriminators. The sub-discriminators obtain different periodic parts of raw waveforms. The method of HiFiSinger proposed by Chen \textit{et al.}~\cite{chen2020hifisinger}, multi-scale adversarial training in both the acoustic model and vocoder was introduced to tackle the difficulty of singing modeling caused by the high sampling rate. A difference between singing voice synthesis and speech synthesis is that the prosody information in the song is more complex~\cite{lee2019adversarially,gu2021bytesing,zhang2021singer,asru2021tang}. The vocal mechanism of singing and voice is different, and the pitch is relatively stable in singing. But the duration of each word is dynamic change according to the music score. We need to pay more attention to the pitch and duration of phonemes. The input of the model in singing voice synthesis often consists of three parts, namely, the lyrics, the F0 corresponding to the lyrics pitch, and the duration corresponding to the lyrics~\cite{BonadaUB16,zhang2022Singer}.

The singing voice synthesis method has experienced the traditional concatenative method based on lyrics to singing alignment~\cite{chien2016alignment,asru2021zhang,gupta2018semi}, and the synthesis method based on statistical parameters~\cite{blaauw2017neural,aolan2021}, to the present end-to-end deep learning method~\cite{hono2019singing,sibo2022,nakamura2019singing,zhang2022MetaSID,yi2019singing}. The concatenative based synthesis method has high sound quality but weak generalization ability and phonemes that are not covered in the training set cannot be generated~\cite{kim2018korean}. The synthesis method based on statistical parameters can solve the generalization problem outside the training set, but the synthesis method based on parameters often leads to an over-smoothing effect, making the synthesized singing voice less naturally~\cite{ChandnaBBG19,zhang2022MDCNN-SID}. The end-to-end synthesis method can learn the mapping relationship between the input music score or lyrics and the acoustic features. 

Many methods also prove that the end-to-end singing synthesis based on the deep model can improve the quality of singing voice. Nakamura \textit{et al.}~\cite{Nakamura20icassp_fast} used CNN for the singing voice synthesis method and the CNN was used to learn the temporal relationship between frame segments of the singing voice. Ren \textit{et al.}~\cite{Ren20kdd_deepsinger} used a forward transformer-based network to perform end-to-end singing voice synthesis, directly generating a linear spectrum and then obtaining the singing voice through the Griffin-Lim vocoder. Shi \textit{et al.}~\cite{shi20arxiv_sequence} combined the perceptual entropy loss function with mainstream time sequence models, including RNN, transformer, and conformer for singing voice synthesis. Xue \textit{et al.}~\cite{Xue20arxiv_learn2sing} used an acoustic model of the encoder-decoder architecture to perform end-to-end training on frame-level input. In the decoder, the RNN uses the current encoder output and the Mel spectrum of the previous time sequence as input to predict the Mel spectrum of the current time sequence. Lu \textit{et al.}~\cite{Lu20Interspeech_xiaoicesing} added F0 prediction and duration prediction for singing voice synthesis based on phoneme modeling based on the FastSpeech model in TTS. Choi \textit{et al.}~\cite{Choi20icassp_korean} used a GAN-based singing synthesis method combined with boundary balance as the objective function and used an autoregressive method to generate a spectrum to solve the time continuity between two adjacent outputs. 

In this paper, we propose an end-to-end singing voice synthesis method with a Striped U-net (SU-net). The proposed method is motivated by the successfully adapted U-net~\cite{Jansson17ISMIR_singing} on the singing voice separation task, which maps the spectrum to the mask of the target spectrum. In the proposed method, the singing voice synthesis task is regarded as the translation task from the encoded embedding of the musical score to the spectrum. Regarding the relationship between the fundamental frequency and the harmonics composition of the singing voice in the spectrum. We use stripe pooling during the convolution process to draw attention to frequency bins dependence on time and the composition relationship on the frequency domain. 

Our contributions are as follows:
\begin{itemize}
 \item To alleviate the complexity of the encoder of the music score, we proposed to use the SU-net architecture model in the end to end way for the translation between music score embedding to audio spectrum.
 \item We introduced the stripe pooling module in our proposed method to enhance the learning of the pitch and harmonics information in the spectrum.
\end{itemize}

\section{Related Works}
\subsection{Singing Voice Synthesis}
Singing synthesis is a similar task to speech synthesis, with a number of successful applications in pop music and music production. SVS systems generate the singing voice from a musical score. Previous works have conducted studies on SVS from different aspects, including acoustic modeling~\cite{Nishimura2016SingingVS}, parametric synthesis~\cite{kim2018korean,zhang2022TDASS}, lyrics-to-singing alignment~\cite{7523215}, and adversarial synthesis~\cite{8683154}.

With the successful application of the deep model on many task~\cite{csmt2021sun,zhao2022nnspeech,wang2022drvc,tang2022avqvc,wang2020transfer}, many deep model methods were tried on SVS. Kim \textit{et al.}~\cite{kim2018korean} proposed a Korean SVS system based on a long-short term memory recurrent neural network. Based on the Korean syllable structure, they tried a feature composing method and adopt LSTM for the SVS.

To tackle the difficulty of singing modeling caused by the high sampling rate, Chen \textit{et al.}~\cite{chen2020hifisinger} proposed HiFiSinger, an SVS system towards high-fidelity singing voice using 48kHz sampling rate. They introduced multi-scale adversarial training in both the acoustic model and vocoder to enhance singing modeling.

Nakamura~\cite{nakamura2019singing} used convolution neural networks (CNNs) for SVS, through CNN to model the long-term dependencies of singing voices. The learned acoustic feature sequence is generated for each segment that covers long-term frames. 

Expect the application of deep models on SVS, there are also some studies that were transferred from TTS. Lu \textit{et al.}~\cite{Lu20Interspeech_xiaoicesing} proposed the XiaoiceSing, which follows the main architecture of FastSpeech. While proposing some singing-specific design, it contains features from the musical score (e.g. note pitch and length).

To address the problem of the small amount of training data, Ren \textit{et al.}~\cite{Ren20kdd_deepsinger} developed DeepSinger, which is a multi-lingual multi-singer singing voice synthesis system. It was built from scratch using singing training data mined from music websites.

GANs have been adapted for TTS in recent years. Chandna \textit{et al.}~\cite{ChandnaBBG19} proposed WGANSing, it used vocoder parameters for acoustic modeling, to separate the influence of pitch and timbre. It facilitates the modeling of the large variability of the pitch in the singing voice. 
\subsection{Baseline Method}

The CNN based method of singing voice synthesis from~\cite{Nakamura20icassp_fast}, named as CNNSing, model the long-term dependencies of singing voices. An acoustic feature sequence is generated for each segment that consists of long-term frames, and a natural trajectory is obtained without the parameter generation algorithm.  The acoustic feature was converted to singing voice waveform audio by a WaveNet vocoder.

The U-net based method of singing voice synthesis from \cite{ChandnaBBG19}, named as WGANSing, used the vocoder parameters for acoustic modeling. Wasserstein-GAN model was adapted for singing voice synthesis, the U-net architecture was used in the generator. Finally, the WORLD vocoder was used for acoustic modeling of the singing voice.

\section{Proposed Method}

\begin{figure}[!ht]
	\centering
	\includegraphics[width=\linewidth]{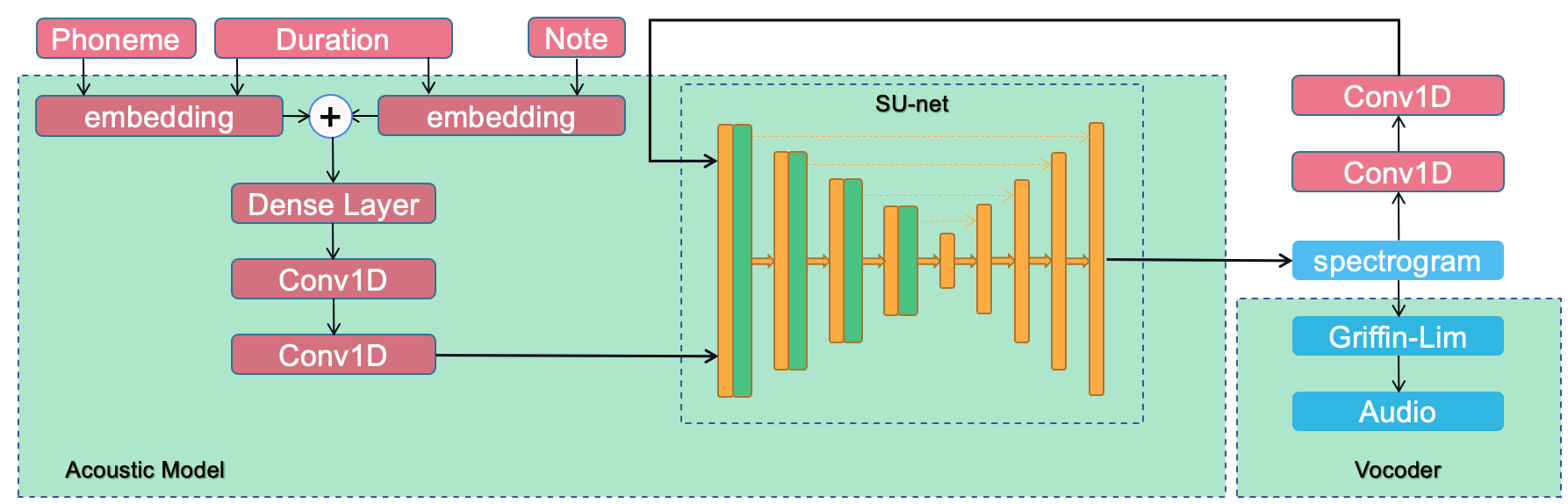}
	\caption{An overview of the proposed method of SU-net for singing voice synthesis. The yellow bar represents convolution layers and deconvolution layers, the green bar represents stripe pooling layer.}
	\label{fig:overview}
\end{figure}

The singing voice synthesis system generally consists of two main modules. Namely, the acoustic model encodes the input musical score and lyrics into acoustic features, and the vocoder generates audio waveform data from the acoustic features. Figure~\ref{fig:overview} shows the overall architecture of the proposed SU-net based singing synthesis system.

\subsection{Musical Score Encoder}

Our singing voice synthesis system uses music scores and lyrics phonemes as input. The music score contains the beginning and end of each note. The note's value is used to determine whether the note is pronounced. Moreover, the onset of all notes is obtained. Lyrics are transformed into 34 phonemes according to the composition of Japanese pronunciation. The labels of the lyrics in the dataset record the corresponding start and end times are in the audio and phonemes units. The representation of phonemes is numerically coded according to the corresponding positions in the complete set of phonemes to obtain all information with the onset and offset.

The frame rate is obtained according to the audio sampling rate and the frame size, and the time stamp sequence of the note and the phoneme are converted to the frame sequence according to the frame rate, and the shortest duration between the music score and the lyrics is used as the aligned embedding length. The encoded note and phoneme are concatenated as the input of the dense layer. Then through two layers of convolution to form the pre-net, the hidden layer output is used as the feature representation of the music score and lyrics.

\subsection{SU-net Architecture Convolution}

In this paper, SU-net is proposed to the task of singing voice synthesis. In the SU-net architecture, several convolution layers with stripe pooling layers and deconvolution layers are stacked. Downsampling in each layer of the convolution layer halves the size and doubles the channel, while the reverse is true in the deconvolution layer. After each convolution layer does a process of stripe pooling layer to learn the stripe information. The low-level high-sampling rate information flows directly to the high-sampling output layer through skip-connections between the convolution layer and the deconvolution layer of the same size.

The two ends of SU-net correspond to the feature representations of music scores and lyrics and the corresponding audio spectrum for singing voice synthesis tasks. To enhance the dependency learning on the spectrum sequence, the previous spectrum is added to the SU-net input. The training phase directly uses the previous segment of the spectrum in the ground truth as input and uses the previous segment of the predicted spectrum in the inference stage. The significant difference between the encoding of music scores and spectrum is not suitable for direct concatenate as SU-net input. Therefore, the input spectrum is first encoded by the convolution layer to a size adapted to the SU-net input. After obtaining the predicted spectrum, the vocoder Griffin-Lim converts it into the corresponding audio.

\begin{figure}[t]
	\centering
	\includegraphics[width=1\linewidth]{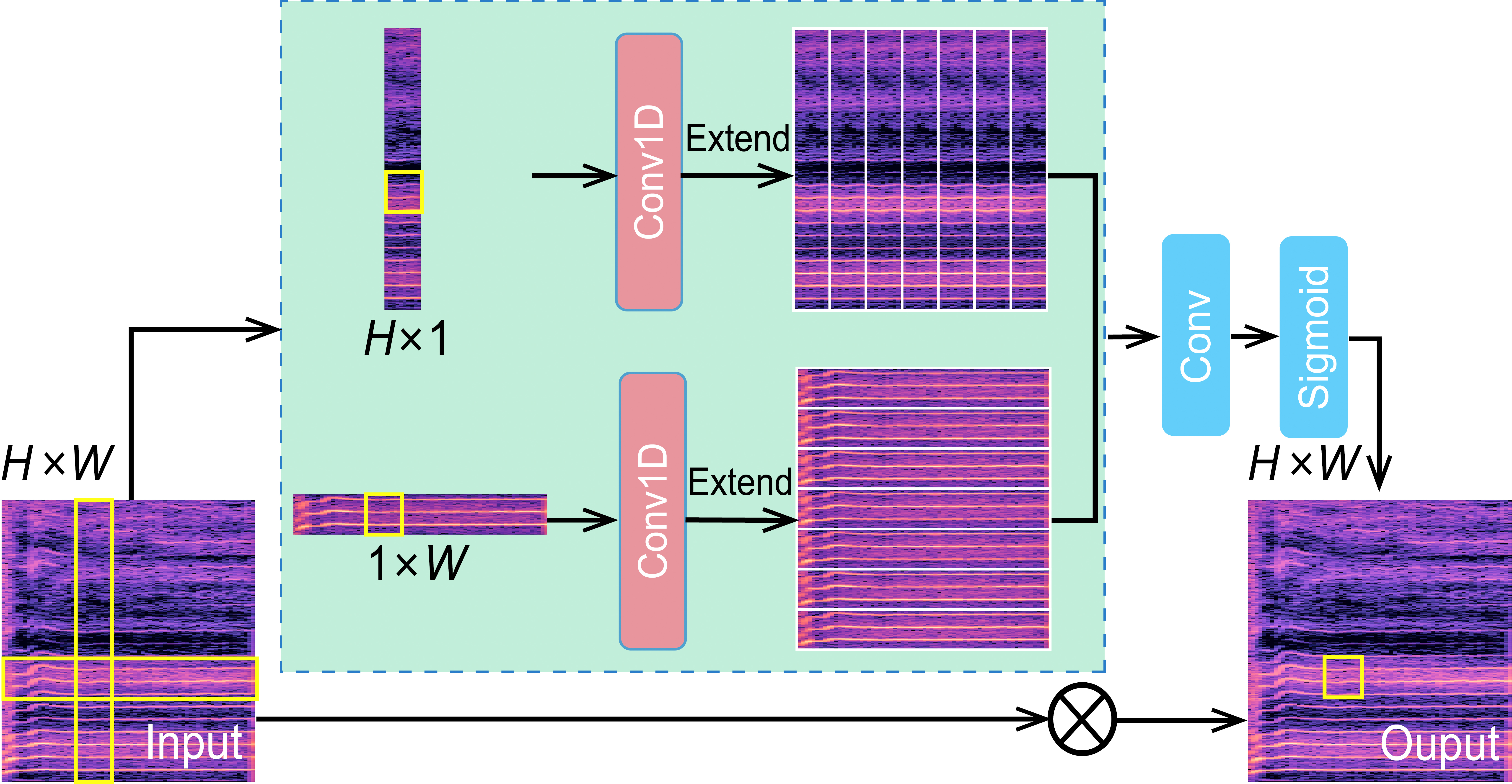}
	\caption{Stripe pooling of the spectrum.}
	\label{fig:stripe pooling}
\end{figure}

To effectively capture the long-term dependence of the fundamental frequency of the singing voice and the harmonics in the time dimension, a method of stripe pooling~\cite{Hou20CVPR_strip} combining horizontal and vertical stripes is append with the convolution layer in SU-net. As shown in Figure~\ref{fig:stripe pooling}, as an alternative to global pooling, stripe pooling has two advantages: firstly, it deploys a long stripe of pooling kernel shape along the horizontal axis time dimension so that it can capture long distances in isolated areas relationship. Secondly, maintaining a narrow stripe of pooling kernel shape in the vertical frequency composition dimension is convenient for capturing the relationship between the local fundamental frequency and the harmonic composition and preventing unrelated regional frequency band interference. Integrating this long and narrow pool kernel enables SU-net networks to aggregate global and local frequency band contexts simultaneously. This is essentially different from the traditional pooling of collecting context from a fixed square area.

For the input spectrum or feature map, the horizontal and vertical stripes are pooled, respectively. Mathematically, given the two-dimensional input $ x\in \mathbb{R}^{H\times W} $, a spatial extend of pooling $(H,1)$ or $(1,W)$ is required. The pooling result is subjected to a layer of one-dimensional convolution and expanded in the horizontal and vertical directions. Thus, the output $y^{h}\in \mathbb{R}^{H}$ can be calculated as

\begin{equation}
	y_i^h=\frac{1}{W}\sum_{0\leq i < W} x_{i,j}.
	\label{eq1}
\end{equation}

The output $y^{w}\in \mathbb{R}^{W}$ can be written as

\begin{equation}
	y_j^v=\frac{1}{H}\sum_{0\leq j < H} x_{i,j}.
	\label{eq2}
\end{equation}

\begin{equation}
	y_{c,i,j}=y_{c,i}^h+y_{c,j}^v.
	\label{eq3}
\end{equation}

The results of the horizontal $y^h$ in Equation (\ref{eq1}) and vertical $y^v$ in Equation (\ref{eq2}) expansion are combined in Equation (\ref{eq3}), where $c$ is the channel of the input tensor. Finally, the input frequency spectrum or feature map $x\in \mathbb{R}^{C\times H\times W}$ is dot-multiplied to obtain the output result in Equation(\ref{eq4}).

\begin{equation}
	\bm{z}=Scale(\bm{x},\sigma(f(\bm{y}))),
	\label{eq4}
\end{equation}

\noindent where $f$ is a convolution with kernel of $(1, 1)$ and $\sigma$ is the sigmoid processing, the $Scale(.,.)$ refers to dot-multiply function.
\section{Experiment}

\subsection{Dataset}

Our experiment uses the public dataset of the Kiritan~\cite{kiritan} singing database, which contains 50 Japanese songs sung by the same female singer. The entire dataset's singing part is about 58 minutes in total, and the audio sampling rate is 96kHz. We resampled the experimental audio data in the data preprocessing and adjusted the sampling rate to 22.05kHz. In the dataset, 45 songs were selected as the training set, and the other five songs were used as the test set.

\subsection{Experiment Settings}

 In the acoustic model, the embedding layer encodes the note and lyrics phonemes respectively and then concatenates them together. The embedded feature vector of the phoneme is 256, the embedded feature vector of the embedded note is 32, and the final output feature vector of the embedded layer is 288. The previous segment spectrum and embedding features are concatenated as SU-net input after passing through two convolution layers. The convolution layer uses the same Conv1D configuration, the input and output sizes are 513, the kernel size is 5, the stride is 1, and the padding is 2. The SU-net network layer stack structure consists of 7 layers of down-sampling convolution layers, 7 layers of stripe pooling layers, and 7-layers up-sampling deconvolution layers. Each layer uses two-dimensional convolution and two-dimensional deconvolution. The kernel size is 5, the stride is 2, padding is 2, dilation is 1. The stripe pooling module performs horizontal pattern pooling and vertical pattern pooling respectively after each convolution layer, and the pooling adopts averaging.


We constructed two baseline systems of CNN~\cite{Nakamura20icassp_fast} and U-net~\cite{ChandnaBBG19} for comparison to evaluate the performance of our proposed method. 
To demonstrate the performance of the proposed singing voice synthesis method, we conduct objective evaluations and subjective evaluations shown in section~\ref{objective evaluation} and section~\ref{subjective evaluation}.

\subsection{Objective Evaluations}
\label{objective evaluation}

\begin{figure}[t]
	\centering
	\includegraphics[width=1\linewidth]{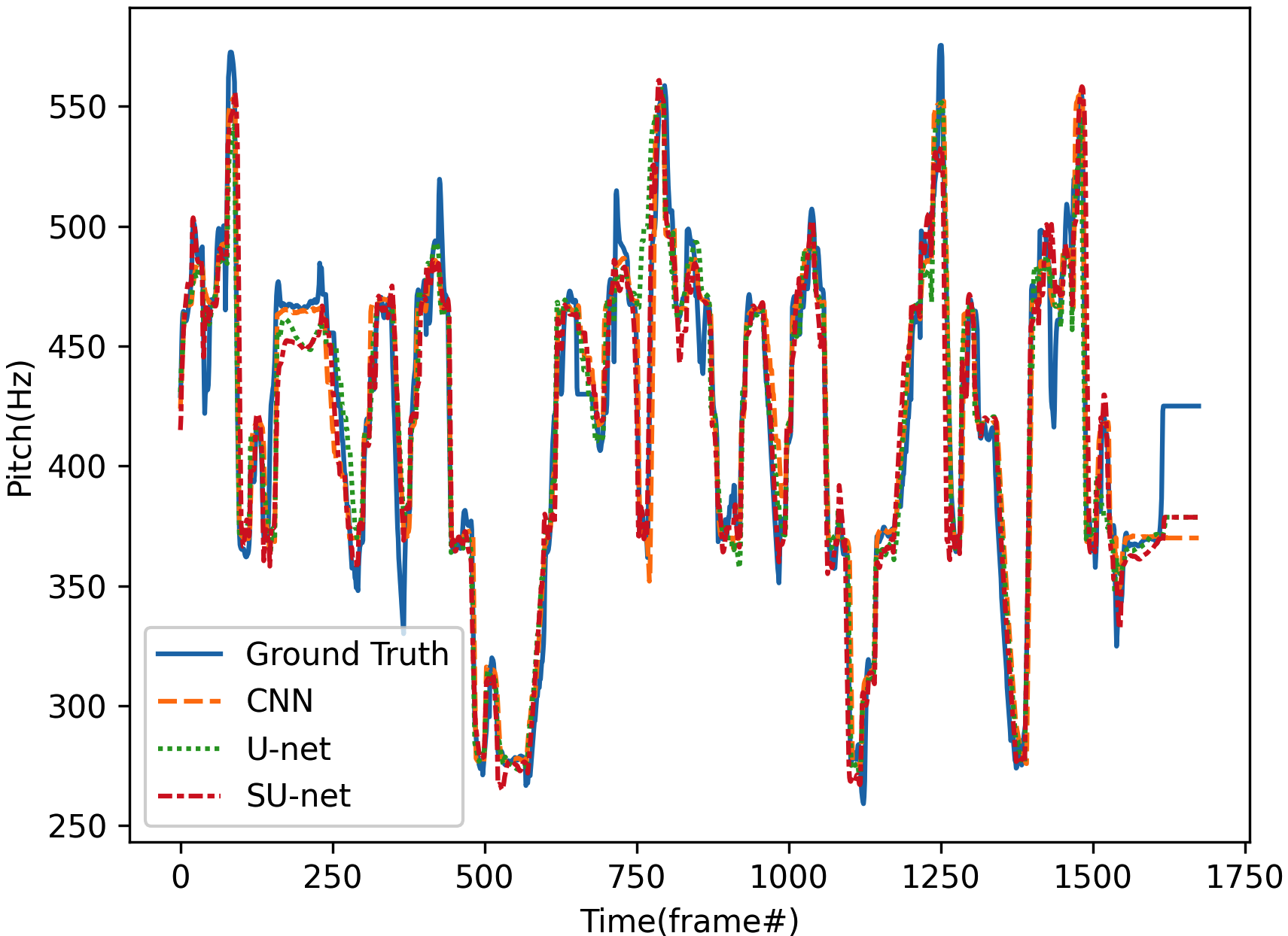}
	\caption{The comparison of F0 between synthetic songs and ground truth.}
	\label{fig: com f0}
\end{figure}

In the objective experiment, F0, voiced and unvoiced segments, and spectrum were compared separately. The F0 comparison is depicted in Figure~\ref{fig: com f0}. The estimation of F0 uses the pYin~\cite{mauch2014pyin} algorithm.

It can be seen from Figure~\ref{fig: com f0}, the consistency of the overall F0 estimation is relatively high, and the F0 variability of the synthetic song is slight in the local details. When compared with the baseline method of CNN and U-net, the estimated pitch nearly have the same performance. These results can conclude that the three methods can well learn the generation of F0. Usually, singers will add a certain amount of vibrato and tremolo according to note pitch according to music score and personal singing skills to make the song sound more varied. On the other hand, the synthetic song tends to be close to a mean when dealing with changes in those frequencies. From the comparison of the pitch estimation, the proposed method can learn the note information. And the pitch information is controllable by changing the input note series. In the experiment, we further tried to change the key of the notes, and the generated singing voice can precisely rise and fall following the key. 

\begin{figure}[t]
	\centering
	\includegraphics[width=1\linewidth]{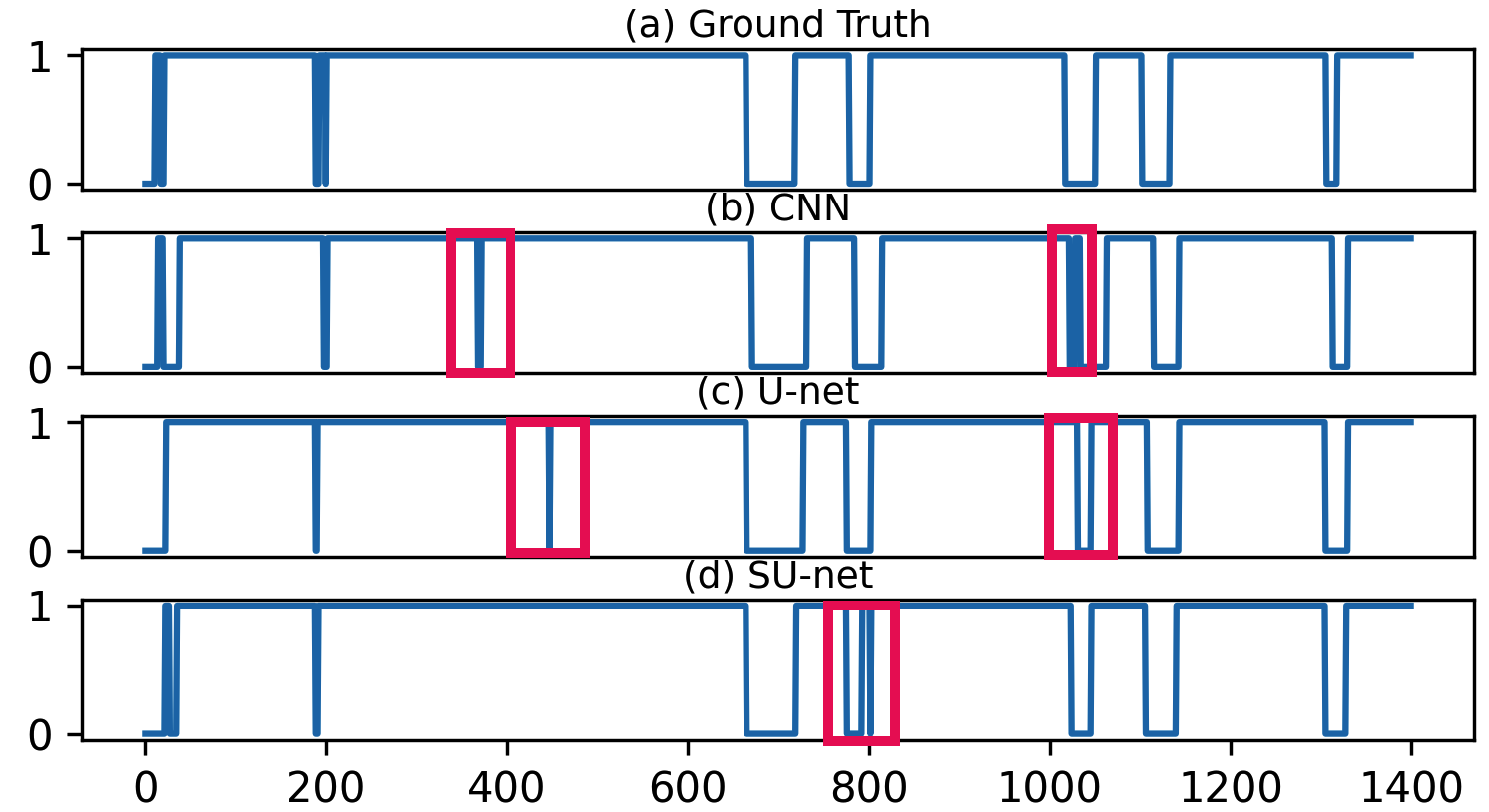}
	\caption{Comparison of the consistency of vocal and silent segments in synthetic songs and ground truth in the time domain.}
	\label{fig: com vocal non vocal}
\end{figure}

In synthetic songs, the vocal and nonvocal parts can be treated as an evaluation index. As shown in Figure~\ref{fig: com vocal non vocal}, it is a comparison of the consistency of the vocal synthesis according to the music score and the voiced and unvoiced segments in the ground truth of the singing voice in the time domain. The detection of voiced and unvoiced segments is acquired by vocal detection using a mute threshold. 

As can be seen from Figure~\ref{fig: com vocal non vocal}, the top subfigure (a) is the ground truth, and the bottom three are the synthetic songs. With vocal detection, the voice frame is marked as 1, and the unvoiced frame is marked as 0. It can be seen that the predicted singing and non-singing of the proposed method based SU-net in the time domain are consistent with the ground truth, and short silent areas are error synthesized nearby frame 800. It could be eliminated by a post-process of temporal smoothing on the vocal detection. While the baseline methods of CNN and U-net have several errors about vocal are synthesized as nonvocal as depicted in the red rectangles. It will lead to a break in the singing voice and affect the continuity of singing.
\begin{figure}[t]
	\centering
	\includegraphics[width=1\linewidth]{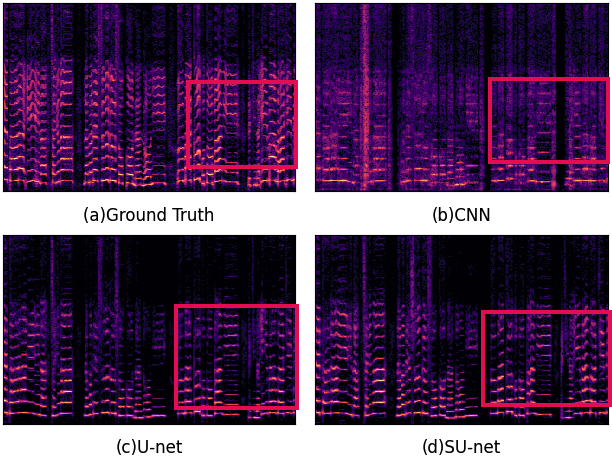}
	\caption{The Mel spectrum of synthetic songs and target song.}
	\label{fig:compare mel spec}
\end{figure}

Furthermore, we inspect the generated spectrogram. The comparison results are shown in Figure~\ref{fig:compare mel spec}. Firstly, we can find that all the synthetic songs have lower energy than the ground truth in high-frequency bands. It shows that some frequency details are discarded of the high-frequency band. When compared to the different methods in low-frequency bands, our proposed method have superior to the baseline methods. As depicted in the red rectangle in the subfigure of Figure~\ref{fig:compare mel spec}, the lines of frequency bins from CNN based synthetic are blurred, while the U-net based and SU-net based are more clearly, and the SU-net has most details close to the ground truth. 

Furthermore, we used Mel-Cepstral Distortion (MCD) as a total statistic metric on the test dataset. The comparison result of MCD is shown in Table~\ref{tab:mcd}.

\begin{table}[]
    \centering
    \caption{The MCD metric for the five songs of the test dataset used for validation of the model. The lower value of MCD, the better.}
    \begin{tabular}{cccc}
    \toprule
         Song & CNNSing~\cite{Nakamura20icassp_fast} & WGANSing~\cite{ChandnaBBG19} & SUSing  \\
         \midrule
         Song 46&9.61&8.73&8.37\\
         Song 47&9.29&8.61&8.38\\
         Song 48&9.82&8.85&8.52\\
         Song 49&9.91&8.92&8.63\\
         Song 50&9.36&8.67&8.42\\ 
         \bottomrule
    \end{tabular}
    \label{tab:mcd}
\end{table}
From the comparison result of the MCD metric, the proposed method of SUSing outperforms the baseline methods of CNNSing and WGANSing. The SUSing has a slight improvement over the WGANSing in terms of MCD, and gets an improvement of 1 over the baseline method of CNNSing. Finally, the proposed method achieved a best MCD score of 8.42 on the test song of Song 50. It shows that the add a stripe module to the U-net architecture will improve the learning of the spectrum.

\subsection{Subjective Evaluations}
\label{subjective evaluation}

The subjective test scores Mean Opinion Score (MOS) from the sound quality, and naturalness. For the singing voice test, 20 segments of 10 seconds were selected for each model, and 10 subjects were given a score of 1-5. Finally, the scoring results of all subjects in the sample are averaged as the final scoring result. 
\begin{table}[th]
	\caption{Mean Opinion Score (MOS) ratings on a 1–5 scale with 95\% confidence intervals.}
	\label{tab:mos}
	\centering
	\begin{tabular}{cccc}
		\toprule
		\multicolumn{1}{c}{\textbf{Model}} &
		\multicolumn{1}{c}{\textbf{MOS}} \\
		\midrule
	Ground Truth & 4.60$\pm$0.12 \\	CNNSing~\cite{Nakamura20icassp_fast} & 3.98$\pm$0.13 \\
		WGANSing~\cite{ChandnaBBG19} & 4.05 $\pm$0.14 \\
		SUSing & 4.12 $\pm$0.15 \\
		
		\bottomrule
	\end{tabular}

\end{table}

The MOS results are shown in Table~\ref{tab:mos}. From the results, it can be seen that the MOS of the synthesized singing voice achieved 4.12. Compared with the baseline methods of CNNSing~\cite{Nakamura20icassp_fast} and WGANSing~\cite{ChandnaBBG19}. The MOS of the U-net based is higher than the baseline CNN, about 0.07. It shows that the U-net architecture convolution improves the use of the low layer of the so-called high-resolution information. The input of the deconvolution layer not only feed with the adjacent layer output but also get the output with the same shape low layer convolution output.

Furthermore, the comparison between WGANSing and SUSing can conclude that the stripe pooling module worked. The Stripe pooling module can utilize the information of horizontal stripes and vertical stripes. While the horizontal stripes in the singing voice spectrum are pitch and harmonics. The striped pooling enhances the pitch and homophones to synthesize a more stable and harmonic singing voice.

\subsection{Ablation Study}

\begin{figure}
    \centering
    \includegraphics[width=1\linewidth]{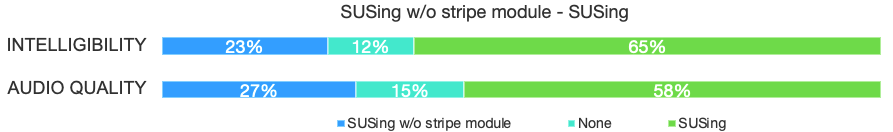}
    \caption{Subjective ABX test for the pair of SUSing and SUSing w/o stripe module.}
    \label{fig:ablation stripe}
\end{figure}

\begin{figure}
    \centering
    \includegraphics[width=1\linewidth]{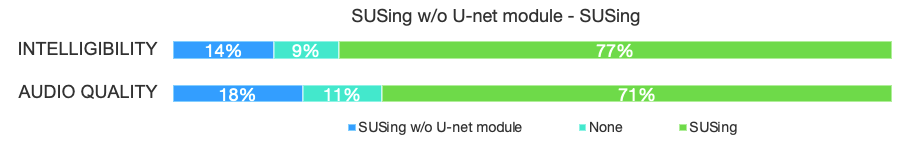}
    \caption{Subjective ABX test for the pair of SUSing and SUSing w/o U-net module.}
    \label{fig:ablation unet}
\end{figure}

To validate the different modules of the proposed method. We further do an ablation study. In the ablation study, we used ABX test, the tester was asked to choose one audio or none in the compare pairs based on the criteria of audio quality and intelligibility. We compared 2 pairs of the model for this evaluation:SUSing w/o stripe module with SUSing, SUSing w/o U-net module with SUSing. The results of the test are shown in Figure~\ref{fig:ablation stripe} and Figure~\ref{fig:ablation unet}.

From the results of ABX test shown in Figure \ref{fig:ablation stripe} and \ref{fig:ablation unet}, it can be seen that our model is qualitatively competitive with regards to both models in the ablation study. This result is supported by the objective measures of MCD in the comparison with the CNN based method of CNNSing and the U-net based methdod of WGANSing in Table~\ref{tab:mcd}. The CNNSing can represent SUSing w/o U-net module in the ablation study, and WGANSing can represent the model of SUSing w/o stripe module.

\section{Conclusion}

We proposed SU-net for singing voice synthesis named SUSing. The proposed method utilizes the low layer high dimension information flow to high layers of U-net to enhance the learning of temporal and spatial relationships in the spectrum of the singing voice. A stripe pooling module was added to the U-net architecture to learn the stripe information of pitch and harmonics in the spectrum structure. The proposed method was implemented on Kiritan, which is a public dataset of Japanese songs. SU-net was proposed to do the translation from the embedding of music score and lyrics to audio spectrum in an end-to-end way. We showed the proposed method is superior to the baselines of CNN based and the non-stripe U-net based method. 
\section{Acknowledgement}
This paper is supported by the Key Research and Development Program of Guangdong Province under grant No.2021B0101400003. Corresponding author is Jianzong Wang from Ping An Technology (Shenzhen) Co., Ltd (jzwang@188.com).
\bibliographystyle{IEEEtran}
\bibliography{mybib}

\begin{thebibliography}{10}
\providecommand{\url}[1]{#1}
\csname url@samestyle\endcsname
\providecommand{\newblock}{\relax}
\providecommand{\bibinfo}[2]{#2}
\providecommand{\BIBentrySTDinterwordspacing}{\spaceskip=0pt\relax}
\providecommand{\BIBentryALTinterwordstretchfactor}{4}
\providecommand{\BIBentryALTinterwordspacing}{\spaceskip=\fontdimen2\font plus
\BIBentryALTinterwordstretchfactor\fontdimen3\font minus
  \fontdimen4\font\relax}
\providecommand{\BIBforeignlanguage}[2]{{%
\expandafter\ifx\csname l@#1\endcsname\relax
\typeout{** WARNING: IEEEtran.bst: No hyphenation pattern has been}%
\typeout{** loaded for the language `#1'. Using the pattern for}%
\typeout{** the default language instead.}%
\else
\language=\csname l@#1\endcsname
\fi
#2}}
\providecommand{\BIBdecl}{\relax}
\BIBdecl

\bibitem{yi2019singing}
Y.-H. Yi, Y.~Ai, Z.-H. Ling, and L.-R. Dai, ``Singing voice synthesis using
  deep autoregressive neural networks for acoustic modeling,'' in \emph{20th
  Annual Conference of the International Speech Communication Association},
  2019.

\bibitem{zhang2020research}
X.~Zhang, Y.~Yu, Y.~Gao, X.~Chen, and W.~Li, ``Research on singing voice
  detection based on a long-term recurrent convolutional network with vocal
  separation and temporal smoothing,'' \emph{Electronics}, vol.~9, no.~9, p.
  1458, 9 2020.

\bibitem{chen2020hifisinger}
J.~Chen, X.~Tan, J.~Luan, T.~Qin, and T.-Y. Liu, ``Hifisinger: Towards
  high-fidelity neural singing voice synthesis,'' \emph{arXiv:2009.01776},
  2020.

\bibitem{hono2021sinsy}
Y.~Hono, K.~Hashimoto, K.~Oura, Y.~Nankaku, and K.~Tokuda, ``Sinsy: A deep
  neural network-based singing voice synthesis system,'' \emph{IEEE/ACM
  Transactions on Audio, Speech, and Language Processing}, vol.~29, pp.
  2803--2815, 2021.

\bibitem{qubo2021}
X.~Qu, J.~Wang, and J.~Xiao, ``Enhancing data-free adversarial distillation
  with activation regularization and virtual interpolation,'' in \emph{{IEEE}
  International Conference on Acoustics, Speech and Signal Processing}.\hskip
  1em plus 0.5em minus 0.4em\relax {IEEE}, 2021, pp. 3340--3344.

\bibitem{xue2021learn2sing}
H.~Xue, S.~Yang, Y.~Lei, L.~Xie, and X.~Li, ``Learn2sing: Target speaker
  singing voice synthesis by learning from a singing teacher,'' in \emph{2021
  IEEE Spoken Language Technology Workshop (SLT)}.\hskip 1em plus 0.5em minus
  0.4em\relax IEEE, 2021, pp. 522--529.

\bibitem{gao2021vocal}
Y.~Gao, X.~Zhang, and W.~Li, ``Vocal melody extraction via hrnet-based singing
  voice separation and encoder-decoder-based f0 estimation,''
  \emph{Electronics}, vol.~10, no.~3, p. 298, 2021.

\bibitem{kong2020hifi}
J.~Kong, J.~Kim, and J.~Bae, ``Hifi-gan: Generative adversarial networks for
  efficient and high fidelity speech synthesis,'' \emph{Advances in Neural
  Information Processing Systems}, vol.~33, pp. 17\,022--17\,033, 2020.

\bibitem{lee2019adversarially}
J.~Lee, H.-S. Choi, and et~al., ``Adversarially trained end-to-end korean
  singing voice synthesis system,'' in \emph{20th Annual Conference of the
  International Speech Communication Association}, 2019.

\bibitem{gu2021bytesing}
Y.~Gu, X.~Yin, Y.~Rao, Y.~Wan, B.~Tang, Y.~Zhang, J.~Chen, Y.~Wang, and Z.~Ma,
  ``Bytesing: A chinese singing voice synthesis system using duration allocated
  encoder-decoder acoustic models and wavernn vocoders,'' in \emph{12th
  International Symposium on Chinese Spoken Language Processing}.\hskip 1em
  plus 0.5em minus 0.4em\relax {IEEE}, 2021, pp. 1--5.

\bibitem{zhang2021singer}
X.~Zhang, J.~Qian, Y.~Yu, Y.~Sun, and W.~Li, ``Singer identification using deep
  timbre feature learning with knn-net,'' in \emph{2021 IEEE International
  Conference on Acoustics, Speech and Signal Processing}.\hskip 1em plus 0.5em
  minus 0.4em\relax IEEE, 2021, pp. 3380--3384.

\bibitem{asru2021tang}
H.~Tang, X.~Zhang, J.~Wang, N.~Cheng, Z.~Zeng, E.~Xiao, and J.~Xiao, ``{TGAVC}:
  Improving autoencoder voice conversion with text-guided and adversarial
  training,'' in \emph{{IEEE} Automatic Speech Recognition and Understanding
  Workshop}.\hskip 1em plus 0.5em minus 0.4em\relax {IEEE}, 2021, pp. 1--6.

\bibitem{BonadaUB16}
J.~Bonada, M.~Umbert, and M.~Blaauw, ``Expressive singing synthesis based on
  unit selection for the singing synthesis challenge 2016,'' in \emph{17th
  Annual Conference of the International Speech Communication
  Association}.\hskip 1em plus 0.5em minus 0.4em\relax {ISCA}, 2016, pp.
  1230--1234.

\bibitem{zhang2022Singer}
X.~Zhang, J.~Wang, N.~Cheng, and J.~Xiao, ``Singer identification for metaverse
  with timbral and middle-level perceptual features,'' in \emph{International
  Joint Conference on Neural Networks, {IJCNN} 2022}.\hskip 1em plus 0.5em
  minus 0.4em\relax {IEEE}, 2022, pp. 1--7.

\bibitem{chien2016alignment}
Y.-R. Chien, H.-M. Wang, and S.-K. Jeng, ``Alignment of lyrics with accompanied
  singing audio based on acoustic-phonetic vowel likelihood modeling,''
  \emph{IEEE/ACM Transactions on Audio, Speech, and Language Processing},
  vol.~24, no.~11, pp. 1998--2008, 2016.

\bibitem{asru2021zhang}
X.~Zhang, J.~Wang, N.~Cheng, E.~Xiao, and J.~Xiao, ``{CycleGEAN}:cycle
  generative enhanced adversarial network for voice conversion,'' in
  \emph{{IEEE} Automatic Speech Recognition and Understanding Workshop
  (ASRU2021)}.\hskip 1em plus 0.5em minus 0.4em\relax {IEEE}, 2021, pp. 1--6.

\bibitem{gupta2018semi}
C.~Gupta, R.~Tong, H.~Li, and Y.~Wang, ``Semi-supervised lyrics and
  solo-singing alignment.'' in \emph{Proceedings of the 19th International
  Society for Music Information Retrieval Conference}, 2018, pp. 600--607.

\bibitem{blaauw2017neural}
M.~Blaauw and J.~Bonada, ``A neural parametric singing synthesizer modeling
  timbre and expression from natural songs,'' \emph{Applied Sciences}, vol.~7,
  no.~12, p. 1313, 2017.

\bibitem{aolan2021}
A.~Sun, J.~Wang, N.~Cheng, M.~Tantrawenith, Z.~Wu, H.~Meng, E.~Xiao, and
  J.~Xiao, ``Reconstructing dual learning for neural voice conversion using
  relatively few samples,'' in \emph{{IEEE} Automatic Speech Recognition and
  Understanding Workshop}.\hskip 1em plus 0.5em minus 0.4em\relax {IEEE}, 2021,
  pp. 946--953.

\bibitem{hono2019singing}
Y.~Hono, K.~Hashimoto, K.~Oura, Y.~Nankaku, and K.~Tokuda, ``Singing voice
  synthesis based on generative adversarial networks,'' in \emph{2019 IEEE
  International Conference on Acoustics, Speech and Signal Processing}.\hskip
  1em plus 0.5em minus 0.4em\relax IEEE, 2019, pp. 6955--6959.

\bibitem{sibo2022}
S.~Si, J.~Wang, J.~Peng, and J.~Xiao, ``Towards speaker age estimation with
  label distribution learning,'' in \emph{ICASSP 2022 - 2022 IEEE International
  Conference on Acoustics, Speech and Signal Processing (ICASSP)}, 2022, pp.
  4618--4622.

\bibitem{nakamura2019singing}
K.~Nakamura, K.~Hashimoto, K.~Oura, Y.~Nankaku, and K.~Tokuda, ``Singing voice
  synthesis based on convolutional neural networks,'' vol. abs/1904.06868,
  2019.

\bibitem{zhang2022MetaSID}
X.~Zhang, J.~Wang, N.~Cheng, and J.~Xiao, ``Metasid: Singer identification with
  domain adaptation for metaverse,'' in \emph{International Joint Conference on
  Neural Networks, {IJCNN} 2022}.\hskip 1em plus 0.5em minus 0.4em\relax
  {IEEE}, 2022, pp. 1--7.

\bibitem{kim2018korean}
J.~Kim, H.~Choi, J.~Park, S.~Kim, J.~Kim, and M.~Hahn, ``Korean singing voice
  synthesis system based on an lstm recurrent neural network,'' in \emph{17th
  Annual Conference of the International Speech Communication Association},
  2018, pp. 1551--1555.

\bibitem{ChandnaBBG19}
P.~Chandna, M.~Blaauw, J.~Bonada, and E.~G{\'{o}}mez, ``Wgansing: {A}
  multi-voice singing voice synthesizer based on the wasserstein-gan,'' in
  \emph{27th European Signal Processing Conference}.\hskip 1em plus 0.5em minus
  0.4em\relax {IEEE}, 2019, pp. 1--5.

\bibitem{zhang2022MDCNN-SID}
X.~Zhang, J.~Wang, N.~Cheng, and J.~Xiao, ``Mdcnn-sid: Multi-scale dilated
  convolution network for singer identification,'' in \emph{International Joint
  Conference on Neural Networks, {IJCNN} 2022}.\hskip 1em plus 0.5em minus
  0.4em\relax {IEEE}, 2022, pp. 1--7.

\bibitem{Nakamura20icassp_fast}
K.~Nakamura, S.~Takaki, K.~Hashimoto, K.~Oura, Y.~Nankaku, and K.~Tokuda,
  ``Fast and high-quality singing voice synthesis system based on convolutional
  neural networks,'' in \emph{2020 {IEEE} International Conference on
  Acoustics, Speech and Signal Processing}.\hskip 1em plus 0.5em minus
  0.4em\relax {IEEE}, 2020, pp. 7239--7243.

\bibitem{Ren20kdd_deepsinger}
Y.~Ren, X.~Tan, T.~Qin, J.~Luan, Z.~Zhao, and T.~Liu, ``Deepsinger: Singing
  voice synthesis with data mined from the web,'' in \emph{26th {ACM} {SIGKDD}
  Conference on Knowledge Discovery and Data Mining}.\hskip 1em plus 0.5em
  minus 0.4em\relax {ACM}, 2020, pp. 1979--1989.

\bibitem{shi20arxiv_sequence}
J.~Shi, S.~Guo, N.~Huo, Y.~Zhang, and Q.~Jin, ``Sequence-to-sequence singing
  voice synthesis with perceptual entropy loss,'' in \emph{2021 IEEE
  International Conference on Acoustics, Speech and Signal Processing}.\hskip
  1em plus 0.5em minus 0.4em\relax {IEEE}, 2021, pp. 76--80.

\bibitem{Xue20arxiv_learn2sing}
H.~Xue, S.~Yang, Y.~Lei, L.~Xie, and X.~Li, ``Learn2sing: Target speaker
  singing voice synthesis by learning from a singing teacher,'' in \emph{2021
  IEEE Spoken Language Technology Workshop}.\hskip 1em plus 0.5em minus
  0.4em\relax {IEEE}, 2021, pp. 522--529.

\bibitem{Lu20Interspeech_xiaoicesing}
P.~Lu, J.~Wu, J.~Luan, X.~Tan, and L.~Zhou, ``Xiaoicesing: {A} high-quality and
  integrated singing voice synthesis system,'' in \emph{21st Annual Conference
  of the International Speech Communication Association}.\hskip 1em plus 0.5em
  minus 0.4em\relax {ISCA}, 2020, pp. 1306--1310.

\bibitem{Choi20icassp_korean}
S.~Choi, W.~Kim, S.~Park, S.~Yong, and J.~Nam, ``Korean singing voice synthesis
  based on auto-regressive boundary equilibrium gan,'' in \emph{2020 {IEEE}
  International Conference on Acoustics, Speech and Signal Processing}.\hskip
  1em plus 0.5em minus 0.4em\relax {IEEE}, 2020, pp. 7234--7238.

\bibitem{Jansson17ISMIR_singing}
A.~Jansson, E.~J. Humphrey, N.~Montecchio, R.~M. Bittner, A.~Kumar, and
  T.~Weyde, ``Singing voice separation with deep u-net convolutional
  networks,'' in \emph{Proceedings of the 18th International Society for Music
  Information Retrieval Conference}, 2017, pp. 745--751.

\bibitem{Nishimura2016SingingVS}
M.~Nishimura, K.~Hashimoto, K.~Oura, Y.~Nankaku, and K.~Tokuda, ``Singing voice
  synthesis based on deep neural networks,'' in \emph{17th Annual Conference of
  the International Speech Communication Association}, 2016.

\bibitem{zhang2022TDASS}
X.~Zhang, J.~Wang, N.~Cheng, and J.~Xiao, ``Tdass: Target domain adaptation
  speech synthesis framework for multi-speaker low-resource tts,'' in
  \emph{International Joint Conference on Neural Networks, {IJCNN} 2022}.\hskip
  1em plus 0.5em minus 0.4em\relax {IEEE}, 2022, pp. 1--7.

\bibitem{7523215}
Y.-R. Chien, H.-M. Wang, and S.-K. Jeng, ``Alignment of lyrics with accompanied
  singing audio based on acoustic-phonetic vowel likelihood modeling,''
  \emph{IEEE/ACM Transactions on Audio, Speech, and Language Processing},
  vol.~24, no.~11, pp. 1998--2008, 2016.

\bibitem{8683154}
Y.~Hono, K.~Hashimoto, K.~Oura, Y.~Nankaku, and K.~Tokuda, ``Singing voice
  synthesis based on generative adversarial networks,'' in \emph{2019 IEEE
  International Conference on Acoustics, Speech and Signal Processing}, 2019,
  pp. 6955--6959.

\bibitem{csmt2021sun}
Y.~Sun, X.~Zhang, X.~Chen, Y.~Yu, and W.~Li, ``Investigation of singing voice
  separation for singing voice detection in polyphonic music,'' in
  \emph{Proceedings of the 9th Conference on Sound and Music Technology}, 2021,
  pp. 1--6.

\bibitem{zhao2022nnspeech}
B.~Zhao, X.~Zhang, J.~Wang, N.~Cheng, and J.~Xiao, ``nnspeech: Speaker-guided
  conditional variational autoencoder for zero-shot multi-speaker
  text-to-speech,'' in \emph{2022 IEEE International Conference on Acoustics,
  Speech and Signal Processing}.\hskip 1em plus 0.5em minus 0.4em\relax IEEE,
  2022, pp. 1--5.

\bibitem{wang2022drvc}
Q.~Wang, X.~Zhang, J.~Wang, N.~Cheng, and J.~Xiao, ``Drvc: A framework of
  any-to-any voice conversion with self-supervised learning,'' in \emph{2022
  IEEE International Conference on Acoustics, Speech and Signal
  Processing}.\hskip 1em plus 0.5em minus 0.4em\relax IEEE, 2022, pp. 1--5.

\bibitem{tang2022avqvc}
H.~Tang, X.~Zhang, J.~Wang, N.~Cheng, and J.~Xiao, ``Avqvc: One-shot voice
  conversion by vector quantization with applying contrastive learning,'' in
  \emph{2022 IEEE International Conference on Acoustics, Speech and Signal
  Processing}.\hskip 1em plus 0.5em minus 0.4em\relax IEEE, 2022, pp. 1--5.

\bibitem{wang2020transfer}
L.~Wang, H.~Zhu, X.~Zhang, S.~Li, and W.~Li, ``Transfer learning for music
  classification and regression tasks using artist tags,'' in \emph{Proceedings
  of the 7th Conference on Sound and Music Technology (CSMT2019)}.\hskip 1em
  plus 0.5em minus 0.4em\relax Springer, 2020, pp. 81--89.

\bibitem{Hou20CVPR_strip}
Q.~Hou, L.~Zhang, M.~Cheng, and J.~Feng, ``Strip pooling: Rethinking spatial
  pooling for scene parsing,'' in \emph{2020 {IEEE/CVF} Conference on Computer
  Vision and Pattern Recognition}.\hskip 1em plus 0.5em minus 0.4em\relax
  {IEEE}, 2020, pp. 4002--4011.

\bibitem{kiritan}
I.~Ogawa and M.~Morise, ``Tohoku kiritan singing database: A singing database
  for statistical parametric singing synthesis using japanese pop songs,''
  \emph{Acoustical Science and Technology}, vol.~42, no.~3, pp. 140--145, 2021.

\bibitem{mauch2014pyin}
M.~{Mauch} and S.~{Dixon}, ``Pyin: A fundamental frequency estimator using
  probabilistic threshold distributions,'' in \emph{2014 IEEE International
  Conference on Acoustics, Speech and Signal Processing}.\hskip 1em plus 0.5em
  minus 0.4em\relax {IEEE}, 2014, pp. 659--663.

\end{thebibliography}

\end{document}